\documentclass{article}%
\usepackage{amsfonts}
\usepackage{amsmath}%
\usepackage{amssymb}%
\usepackage{graphicx}
\providecommand{\U}[1]{\protect\rule{.1in}{.1in}}
\newtheorem{theorem}{Theorem}
\newtheorem{acknowledgement}[theorem]{Acknowledgement}

\begin{document}

\title{Equations for spinning test particles in equatorial orbits when they are
orbiting in a weak rotating field}
\author{Nelson Velandia, and Juan Manuel Tejeiro\\Universidad Nacional de Colombia}
\maketitle

\begin{abstract}
This paper formulates, via the Mathisson - Papapetrou - Dixon equations, the
system of equations for a test particle with spin when it is orbiting a weak
Kerr metric. We shall restrict ourselves to the case of circular orbits with
the purpose of comparing our results with the results of the literature. In
particular, we solve the set of equations of motion for the case of circular
trajectories both spinless and spinning test particles around rotating bodies
in equatorial plane. The results obtained are an important guideline for the
study of the effects of the particles with spin in rotating gravitational
fields such as Gravitomagnetics Effects or gravitational waves.

\end{abstract}

\section{Introduction}

Presently, there exists an interest in the study of the effects of the spin in
the trajectory of test particles in rotating gravitational fields. The
importance of this topic increases when dealing with phenomena of astrophysics
such as accretion discs in rotating black holes \cite{tanaka 1996},
Gravitomagnetics Effects \cite{faruque2004} or gravitational waves induced by
spinning particles orbiting a rotating black hole \cite{mino}. Therefore, we
shall work the equations of motion for test particles in a weak Kerr metric
which will be integrated numerically in the particular case when the test
particles are orbiting circularlly with the purpose of studying the effects of
spin in the trajectories of test particles in rotating gravitational fields.

The motion of particles in a gravitational field is given by the geodesics
equation. The solution to this equation depends on the problem, and therefore
there are different methods for its solution \cite{carter} \cite{abramowicz}.
Basically, we take two cases in motion of test particles in a gravitational
field of a rotating mass. The first case describes the trajectory of a
spinless test particle, and the second one the trajectory of a spinning test
particle in a weak Kerr metric. For the second case a representation is used
that does not include the third-order derivatives of the coordinates, and
yields the equations of motion for a spinning test particle in a gravitational
field without any restrictions on its velocity and spin orientations
\cite{plyatsko}.

For the first case, authors as Tanaka \textit{et al. } \cite{tanaka 1996}
yield the set of equations of motion for orbiting spinless test particles. In
this case the equations of motion for the spinless test particle are
considered both in the equatorial \cite{bardeen}, \cite{wilkins},
\cite{calvani}, and the non-equatorial plane \cite{wilkins}, \cite{stog},
\cite{teo} (Kheng, L., Perng, S., Sze Jackson, T.: Massive Particle Orbits
Around Kerr Black Holes. Unpublished).

For the study of test particles in a rotating field, some authors have solved
the equations of motion for spinless and spinning test particles in the
particular case of circular orbits in the equatorial plane of a Kerr metric
\cite{tanaka 1996}, \cite{bardeen}, \cite{suzuki}, \cite{mash}, \cite{dadhich}%
, \cite{bini 2005}, \cite{tod}, \cite{tartaglia}. In addition, Plyastko, R.
\textit{et. al. }yield the full set of Mathisson-Papapetrou-Dixon Equations
(MPD equations) for spinning test particles in the Kerr gravitational field
\cite{plyatsko}. These authors integrate numerically the MPD equations for the
case of the Schwarzschild metric. In this paper, we use the method of MPD
Equations given by Plyastko, R. \textit{et. al. }for calculating the
trajectories of spinless and spinning test particles in equatorial planes for
circular orbits, i.e., constant radius in a weak Kerr metric. In the
literature, there are not works that study via MPD equations the trajectories
of spining test particles in weak fields.

With the purpose to prove the equations of motion that we worked, we shall
solve numerically the set of equations of motion obtained via MPD Equations in
the case when the spinless test particle is in the equatorial plane and will
compare the results with works that involve astronomy, especially the study of
satellites which orbit around the Earth. We take the same initial conditions
in the two cases for describing the trajectory both a spinless particle and a
spinning particle in a weak Kerr metric. Then, we compare the cartesian
coordinates ($x,y,z$) for the trajectory of two particles that travel in the
same orbit but in opposite directions. We shall take both for a spinless test
particle and for a spinning test particle orbiting in a weak Kerr field.

This work is organized as follows. In Section 2 we give a brief introduction
to the MPD\ Equations that work the set of equations of motion for test
particles both spinless and spinning in a rotating gravitational field. From
the MPD equations of motion, we yield the equations of motion for spinless and
spinning test particles and will study the spinless test particles. Also, we
will give the set of MPD equations given by Plyatsko \textit{et al.
}\cite{plyatsko} in a schematic form for working the case of a weak Kerr
metric. In Section 3, we present the Gravitomagnetic Clock Effect in order to
prove our set of equations for spinless and spinning test particles. Then, in
Section 4, we make a numerical comparison for spinless and spinning test
particle via MPD equations in the equatorial plane. We take the initial values
from a satellite that is orbiting around on the Earth; then, we substitute
these values in the MPD equations both for spinless particles and for spinning
particles, and finally we make a numerical comparison of the trajectory in
cartesian coordinates for two particles that travel in the same orbit, but in
opposite directions. In the last section, the conclusions and some future
works are formulated in order to describe spinning test particles in a weak
Kerr metric.

\section{Brief introduction to the Mathisson-Papapetrou-Dixon Equations}

In general the MPD equations \cite{mathisson}, \cite{papapetrou}, \cite{dixon}
are given by the dynamics of extended bodies in the general theory of
relativity which includes any gravitational background. For the solution of
our problem, we take the case of a distribution of mass ($m$) with a spin
tensor ($S^{\rho\sigma}$) around a rotating central source ($M$) which has a
metric tensor $g_{\mu\nu}$. These equations of motion for a spinning test
particle are obtained in terms of an expansion that depends on the derivatives
of the metric and the multipole moments of the energy-momentum tensor
($T^{\mu\nu}$) \cite{dixon} which describe the motion of an extended body. In
this work, we shall take a body sufficiently small so that all higher
multipoles can be neglected. According to this restriction the MPD equations
are given by%
\begin{equation}
\frac{D}{ds}\left(  mu^{\lambda}+u_{\lambda}\frac{DS^{\lambda\mu}}{ds}\right)
=-\frac{1}{2}u^{\pi}S^{\rho\sigma}R_{\pi\rho\sigma}^{\lambda},\label{mov1}%
\end{equation}

\begin{equation}
\frac{D}{ds}S^{\mu\nu}+u^{\mu}u_{\sigma}\frac{DS^{\nu\sigma}}{ds}-u^{\nu
}u_{\sigma}\frac{DS^{\mu\sigma}}{ds}=0,\label{mov2}%
\end{equation}
where $D/ds$ means the covariant derivative, and the antisymmetric tensor
$S^{\mu\nu}$ are the linear and spin angular momenta, respectively.
$R^{\lambda}{}_{\pi\rho\sigma}$ is the curvature tensor, and $u^{\mu}=dz^{\mu
}/ds$. But we do not have the evolution equation for $u^{\mu}$ and it is
neccesary to single out the center of mass which determines the world line as
a representing path and specifies a point about which the momentum and spin of
the particle are calculated. This world line can be determined from physical
considerations \cite{karpov}. In general, two conditions are usually imposed.
The Mathisson-Pirani supplementary condition is \cite{mathisson} \cite{pirani}%
\begin{equation}
u_{\sigma}S^{\mu\sigma}=0\label{mp}%
\end{equation}
and the Tulczyjew-Dixon condition \cite{dixon}%

\begin{equation}
p_{\sigma}S^{\mu\sigma}=0\label{cond1}%
\end{equation}
where%
\begin{equation}
p^{\sigma}=mu^{\sigma}+u_{\lambda}\frac{DS^{\sigma\lambda}}{ds}%
\end{equation}
is the four momentum.

For to obtain the set of MPD equations, we take the MP condition (\ref{mp})
which has three independent relationships between $S^{\mu\sigma}$ and
$u_{\sigma}$. By this condition $S^{i4}$ is given by%
\begin{equation}
S^{i4}=\frac{u_{k}}{u_{4}}S^{ki}%
\end{equation}
with this expression we can deal the independent components $S^{ik}$.
Sometimes it is more convenient the vector spin which is given by
$\ S_{i}=\frac{1}{2u_{4}}\sqrt{-g}\epsilon_{ikl}S^{kl}$, where $\epsilon
_{ikl}$ is the spatial L\'{e}vi-Civit\`{a} symbol.

On the other hand, when the space-time admits a Killing vector $\xi^{\upsilon
}$, there exists a property that includes the covariant derivative and the
spin tensor, which gives a constant and is given by the expression \cite{dixon
1979}%
\begin{equation}
p^{\nu}\xi_{\nu}+\frac{1}{2}\xi_{\nu,\mu}S^{\nu\mu}=\text{ constant,}%
\label{16}%
\end{equation}
where $p^{\nu}$ is the linear momentum, $\xi_{\nu,\mu}$ is the covariant
derivative of the Killing vector, and $S^{\nu\mu}$ is the spin tensor of the
particle. In the case of the Kerr metric, there are two Killing vectors, owing
to its stationary and axisymmetric nature. In consequence, Eq. (\ref{16})
yields two constants of motion: $E$ is the total energy and $L$ is the
component of its angular momentum along the axis of symmetry \cite{iorio}.

\subsection{MPD Equations for a spinning test particle in a metric of rotating
body}

Given that the spinning body test is sufficiently small in regard to the
characteristic length the equations of motion (Eqs. \ref{mov1} and \ref{mov2})
are reduced to the case when the test particles are orbiting a metric of
rotating body. Then, we will give the equations of motion for the case of a
spinning test particle for a weak Kerr metric (Appendix A).

First of all, we take the paper by R.M. Plyatsko \textit{et al. }%
\cite{plyatsko} for obtaining the full set of the exact MPD equations for the
motion of a spinning test particle in the Kerr field if the MP condition
(\ref{mp}) is taken into account and obtain a general scheme for the set of
equations of motion for a spinning test particle in a rotating field. Plyatsko
\textit{et al. }use the dimensionless quantities \textit{y}$_{i}$ with
particle%
\'{}%
s coordinates by%
\begin{equation}
y_{1}=\frac{r}{M}\text{, \ \ \ \ \ \ }y_{2}=\theta\text{, \ \ \ \ }%
y_{3}=\varphi\text{, \ \ \ \ \ }y_{4}=\frac{t}{M}\label{y1}%
\end{equation}
for its 4-velocity%
\begin{equation}
y_{5}=u^{1}\text{, \ \ \ \ \ }y_{6}=Mu^{2}\text{, \ \ \ }y_{7}=Mu^{3}\text{,
\ \ \ }y_{8}=u^{4}\label{y5}%
\end{equation}
and the spin components \cite{plyatsko 2010}%
\begin{equation}
y_{9}=\frac{S_{1}}{mM}\text{, \ \ \ \ \ \ \ }y_{10}=\frac{S_{2}}{mM^{2}%
}\text{, \ \ \ \ \ }y_{11}=\frac{S_{3}}{mM^{2}}\label{y9}%
\end{equation}

In addition, they introduce another dimensionless quantities with regard to
the proper time $s$ and the constant of motion $E$, $J_{z}$%
\begin{equation}
x=\frac{s}{M}\text{, \ \ \ \ \ }\widehat{E}=\frac{E}{m}\text{, \ \ \ }%
\widehat{J}=\frac{J_{z}}{mM}%
\end{equation}

The set of the MPD equations for a spinning particle in the Kerr field is
given by eleven equations. The first four equations are%
\begin{equation}
\overset{\bullet}{y}_{1}=y_{5}\text{, \ \ \ \ \ }\overset{\bullet}{y}%
_{2}=y_{6}\text{, \ \ \ \ \ }\overset{\bullet}{y}_{3}=y_{7}\text{,
\ \ \ \ }\overset{\bullet}{y}_{4}=y_{8}%
\end{equation}
where a dot denotes the usual derivative with respect to $x$.

The fifth equation is given by the first three equations of (\ref{mov1}) with
the indexes $\lambda=1,2,3$. The result is multiplied by $S_{1,}S_{2},S_{3}$
and with the MP condition (\ref{mp}) we have the relationship: $S^{i4}%
=\frac{u_{k}}{u_{4}}S^{ki}$ and $S_{i}=\frac{1}{2u_{4}}\sqrt{-g}%
\varepsilon_{ikl}S^{kl}$, we obtain%
\begin{equation}
mS_{i}\frac{Du^{i}}{ds}=-\frac{1}{2}u^{\pi}S^{\rho\sigma}S_{j}R_{\pi\rho
\sigma}^{j}%
\end{equation}
which can be written as%
\begin{equation}
y_{9}\overset{\bullet}{y}_{5}+y_{10}\overset{\bullet}{y}_{6}+y_{11}%
\overset{\bullet}{y}_{7}=A-y_{9}Q_{1}-y_{10}Q_{2}-y_{11}Q_{3}%
\end{equation}
where%
\begin{equation}
Q_{i}=\Gamma_{\mu\nu}^{i}u^{\mu}u^{\nu}\text{, \ \ \ \ }A=\frac{u^{\pi}}%
{\sqrt{-g}}u_{4}\epsilon^{i\rho\sigma}S_{i}S_{j}R_{\pi\rho\sigma}^{j}%
\end{equation}

The sixth equation is given by
\begin{equation}
u_{\nu}\frac{Du^{\nu}}{ds}=0
\end{equation}
which can be written as%
\begin{equation}
p_{1}\overset{\bullet}{y}_{5}+p_{2}\overset{\bullet}{y}_{6}+p_{3}%
\overset{\bullet}{y}_{7}+p_{4}\overset{\bullet}{y}_{8}=-p_{1}Q_{1}-p_{2}%
Q_{2}-p_{3}Q_{3}-p_{4}Q_{4}%
\end{equation}
where%
\begin{equation}
p_{\alpha}=u_{\alpha}=g_{\mu\alpha}u^{\alpha}%
\end{equation}

The seventh equation is given by%
\begin{equation}
E=P_{4}-\frac{1}{2}g_{4\mu,\nu}S^{\mu\nu}%
\end{equation}
which can be written as%
\begin{equation}
c_{1}\overset{\bullet}{y}_{5}+c_{2}\overset{\bullet}{y}_{6}+c_{3}%
\overset{\bullet}{y}_{7}=C-c_{1}Q_{1}-c_{2}Q_{2}-c_{3}Q_{3}+\text{\ }%
\widehat{E}%
\end{equation}
where%
\[
d=\frac{1}{\sqrt{-g}}%
\]

\begin{align}
c_{1}  &  =-dg_{11}g_{22}g_{44}u^{2}S_{3}-d\left(  g_{34}^{2}-g_{33}%
g_{44}\right)  g_{11}u^{3}S_{2}\nonumber\\
c_{2}  &  =dg_{11}g_{22}g_{44}u^{1}S_{3}+d\left(  g_{34}^{2}-g_{33}%
g_{44}\right)  g_{22}u^{3}S_{1}\nonumber\\
c_{3}  &  =d\left(  g_{34}^{2}-g_{33}g_{44}\right)  g_{11}u^{1}S_{2}-d\left(
g_{34}^{2}-g_{33}g_{44}\right)  g_{22}u^{2}S_{1}%
\end{align}

\begin{equation}
C=g_{44}u^{4}-dg_{44}u^{4}g_{43,2}S_{1}+d\left(  g_{44}u^{4}g_{43,1}%
-g_{33}u^{3}g_{44,1}\right)  S_{2}+dg_{22}u^{2}g_{44,1}S_{3}%
\end{equation}

The eighth equation is given by%
\begin{equation}
J_{z}=-P_{3}+\frac{1}{2}g_{3\mu,\nu}S^{\mu\nu}%
\end{equation}
which can be written as%
\begin{equation}
d_{1}\overset{\bullet}{y}_{5}+d_{2}\overset{\bullet}{y}_{6}+d_{3}%
\overset{\bullet}{y}_{8}=D-d_{1}Q_{1}-d_{2}Q_{2}-d_{3}Q_{4}-\widehat{J}%
\end{equation}
where%
\begin{align}
d_{1}  &  =-dg_{11}g_{22}g_{34}u^{2}S_{3}+dg_{11}g_{33}g_{34}u^{3}%
S_{2}+dg_{11}g_{34}^{2}u^{4}S_{2}-dg_{11}g_{33}g_{44}u^{4}S_{2}\nonumber\\
d_{2}  &  =-dg_{11}g_{22}g_{34}u^{1}S_{3}-dg_{22}g_{33}g_{34}u^{3}%
S_{1}-dg_{22}g_{34}^{2}u^{4}S_{1}+dg_{22}g_{33}g_{44}u^{4}S_{1}\nonumber\\
d_{3}  &  =-dg_{11}g_{34}^{2}u^{1}S_{2}+dg_{22}g_{34}^{2}u^{2}S_{1}%
+dg_{22}g_{33}g_{44}u^{2}S_{1}-dg_{11}g_{33}g_{34}u^{1}S_{2}%
\end{align}

\begin{align}
D  & =g_{33}u^{3}-dg_{22}u^{2}g_{34,2}S_{1}+d\left(  g_{44}u^{4}%
g_{33,1}+g_{11}u^{1}g_{34,1}-g_{33}u^{3}g_{34,1}\right)  S_{2}\nonumber\\
& -dg_{11}u^{1}g_{34,1}S_{3}%
\end{align}

Finally, the last three equations are given by%
\begin{equation}
u^{4}\overset{\bullet}{S}_{i}+2\left(  \overset{\bullet}{u}_{[4}u_{i]}-u^{\pi
}u_{\rho}\Gamma_{\pi\lbrack4}^{\rho}u_{i]}\right)  S_{k}u^{k}+2S_{n}%
\Gamma_{\pi\lbrack4}^{n}u_{i]}u^{\pi}=0
\end{equation}
which give the derivatives of three components of vector spin
($\overset{\bullet}{S}_{i}$): $\overset{\bullet}{y}_{9}$, $\overset{\bullet
}{y}_{10}$ and $\overset{\bullet}{y}_{11}$.

After achieving the system of equations of motion for spinning test particles,
we numerically solve it. We used the fourth-order Runge Kutta method. First we
take the case when a test particle is orbiting far away from the central
source and is in the equatorial plane. For our numerical calculations, we take
the parameters both of the central source and the test particle such as the
radio, the energy, the angular momentum and the components tangential and
radial of the four-velocity ($u^{\mu}$). We calculate the orbit of a test
particle both spinless and spinning a weak Kerr metric in cartesian
coordinates ($x$, $y$, \thinspace$z$). Then, we make a comparison of the time
that a test particle takes to do a lap in the two cases and give some conclusions.

\subsection{Equations of motion for a spinning test particle in a weak Kerr
metric}

In the last section, we obtained the general scheme for the set of equations
of motion of a spinning test particle in the gravitational field of a rotating
body. Now, we yield the set of equations for the case of a spinning test
particle in the equatorial plane of a weak metric Kerr (Appendix A). This set
of equations is given by%

\begin{equation}
r^{\prime}[s]=\frac{dr}{ds}\text{; \ \ \ }\theta^{\prime}[s]=\frac{d\theta
}{ds}=0\text{; \ \ \ }\varphi^{\prime}[s]=\frac{d\varphi}{ds}\text{;
\ \ \ \ }t%
\acute{}%
[s]=\frac{dt}{ds}%
\end{equation}

\begin{align}
\frac{d^{2}r}{ds^{2}}  &  =\left(  \frac{c_{3}d_{3}}{R_{1}}\right)  \left(
\frac{R_{2}}{c_{3}}+\frac{R_{3}}{d_{3}}+R_{4}\right) \\
\frac{d^{2}\varphi}{ds^{2}}  &  =\left(  \frac{-c_{1}d_{3}}{R_{1}}\right)
\left(  \frac{R_{2}}{c_{3}}+\frac{R_{3}}{d_{3}}+R_{4}\right)  +\frac{\left(
B-c_{1}Q_{1}+\overset{\wedge}{E}\right)  }{c_{3}}-Q_{3}\\
\frac{d^{2}t}{ds^{2}}  &  =\left(  \frac{-d_{1}c_{3}}{R_{1}}\right)  \left(
\frac{R_{2}}{c_{3}}+\frac{R_{3}}{d_{3}}+R_{4}\right)  +\frac{\left(
F-d_{1}Q_{1}-\overset{\wedge}{J}\right)  }{d_{3}}%
\end{align}

\[
z:=\left(  r[s]\right)  ^{2}\text{; \ \ \ \ \ \ \ \ \ \ \ }q:=r[s]\left(
r[s]-2\right)  \text{; \ \ \ }\psi:=\left(  r[s]\right)  ^{2}%
\]

\[
\eta:=3\left(  r[s]\right)  ^{2}\text{; \ \ \ \ }\chi:=\left(  r[s]\right)
^{2}\text{; \ \ \ \ \ \ \ \ \ \ \ \ \ }\xi:=\left(  r[s]\right)  ^{2}%
\]

\[
p:=2\alpha\left(  \sin\left(  \frac{\pi}{2}\right)  \right)  ^{2}%
r[s]\frac{d\varphi}{ds}+\left(  \left(  r[s]\right)  ^{2}-2r[s]\right)
\frac{dt}{ds}\text{; }%
\]

\[
\text{\ }p_{1}:=-\left(  1-\frac{2}{r[s]}\right)  ^{-1}\frac{dr}{ds}\text{;
\ \ }p_{2}:=0;
\]

\[
\text{\ }p_{3}:=\frac{2\alpha\left(  \sin\left(  \frac{\pi}{2}\right)
\right)  ^{2}r[s]\frac{dt}{ds}-\left(  \sin\left(  \frac{\pi}{2}\right)
\right)  ^{2}\left(  r[s]\right)  ^{4}\frac{d\varphi}{ds}}{\left(
r[s]\right)  ^{2}}\text{;}%
\]

\[
p_{4}:=\frac{2\alpha\left(  \sin\left(  \frac{\pi}{2}\right)  \right)
^{2}r[s]\frac{d\varphi}{ds}+\left(  \left(  r[s]\right)  ^{2}-2r[s]\right)
\frac{dt}{ds}}{\left(  r[s]\right)  ^{2}}%
\]

\[
c_{1}:=S_{2}\sin\left(  \frac{\pi}{2}\right)  \frac{d\varphi}{ds}\text{;
\ \ \ \ \ }c_{2}:=0\text{; \ \ \ \ \ }c_{3}:=-S_{2}\sin\left(  \frac{\pi}%
{2}\right)  \frac{dr}{ds}%
\]

\[
d_{1}:=-S_{2}\sin\left(  \frac{\pi}{2}\right)  \frac{dt}{ds}\text{;
\ \ \ \ }d_{2}:=0\text{; \ \ \ \ }d_{3}:=S_{2}\sin\left(  \frac{\pi}%
{2}\right)  \frac{dr}{ds}%
\]

\begin{align*}
Q_{1}  &  :=\frac{\left(  \left(  r[s]-2\right)  -r[s]+2\right)  }{\left(
r[s]\right)  \left(  r[s]-2\right)  +\alpha^{2}}\left(  \frac{dr}{ds}\right)
^{2}-\left(  \sin\left(  \frac{\pi}{2}\right)  \right)  ^{2}\left(
r[s]-2\right)  \left(  \frac{d\varphi}{ds}\right)  ^{2}\\
&  +\frac{\left(  r[s]-2\right)  }{\left(  r[s]\right)  ^{3}}\left(  \frac
{dt}{ds}\right)  ^{2}-\frac{2\alpha\left(  \sin\left(  \frac{\pi}{2}\right)
\right)  ^{2}\left(  r[s]-2\right)  }{\left(  r[s]\right)  ^{3}}\frac
{d\varphi}{ds}\frac{dt}{ds}%
\end{align*}

\begin{align*}
Q_{2}  &  :=0\\
Q_{3}  &  :=\frac{2}{\left(  r[s]\right)  }\frac{dr}{ds}\frac{d\varphi}%
{ds}+\frac{2\alpha}{\left(  r[s]\right)  ^{3}\left(  r[s]-2\right)  }\frac
{dr}{ds}\frac{dt}{ds}\\
Q_{4}  &  :=-\frac{6\alpha}{\left(  r[s]\right)  \left(  r[s]-2\right)  }%
\frac{dr}{ds}\frac{d\varphi}{ds}+\frac{2}{\left(  r[s]\right)  \left(
r[s]-2\right)  }\frac{dr}{ds}\frac{dt}{ds}%
\end{align*}

\[
B:=-\left(  1-\frac{2}{\left(  r[s]\right)  }\right)  \frac{dt}{ds}%
-\frac{2\alpha\left(  \sin\left(  \frac{\pi}{2}\right)  \right)  ^{2}}{\left(
r[s]\right)  ^{2}}\frac{dr}{ds}\frac{d\varphi}{ds}+\frac{S_{2}\sin\left(
\frac{\pi}{2}\right)  }{\left(  r[s]\right)  ^{2}}\frac{d\varphi}{ds}%
-\frac{\alpha S_{2}\sin\left(  \frac{\pi}{2}\right)  }{\left(  r[s]\right)
^{4}}\frac{dt}{ds}%
\]

\begin{align*}
F  &  :=-\frac{3\alpha S_{2}\left(  \sin\left(  \frac{\pi}{2}\right)  \right)
^{3}}{\left(  r[s]\right)  ^{2}}\frac{d\varphi}{ds}-\frac{S_{2}\left(  \left(
r[s]\right)  -2\right)  }{\left(  r[s]\right)  ^{2}}\frac{dt}{ds}\\
&  +\frac{\left(
\begin{array}
[c]{c}%
2\alpha\left(  \sin\left(  \frac{\pi}{2}\right)  \right)  ^{2}\left(
r[s]\right)  ^{2}\frac{dr}{ds}\left(  \frac{d\varphi}{ds}\right)  ^{2}\\
+\left(  r[s]\right)  ^{3}\left(  r[s]-2\right)  \frac{dr}{ds}\frac{dt}%
{ds}-2\alpha\left(  \left(  r[s]\right)  -2\right)  \left(  \frac{dt}%
{ds}\right)  ^{2}%
\end{array}
\right)  \left(  \sin\left(  \frac{\pi}{2}\right)  \right)  ^{2}}%
{2\alpha\left(  \sin\left(  \frac{\pi}{2}\right)  \right)  ^{2}r[s]\frac
{d\varphi}{ds}+\left(  \left(  r[s]\right)  ^{2}-2r[s]\right)  \frac{dt}{ds}}%
\end{align*}

\begin{align*}
R_{1}  &  :=c_{3}d_{3}p_{1}-d_{3}p_{3}c_{1}-p_{4}d_{1}c_{3}\\
R_{2}  &  :=p_{3}\left(  c_{1}Q_{1}-B-\overset{\wedge}{E}\right) \\
R_{3}  &  :=p_{4}\left(  d_{1}Q_{1}-F+\overset{\wedge}{J}\right) \\
R_{4}  &  :=-p_{1}Q_{1}%
\end{align*}

\subsection{MPD Equations for spinless test particle in a weak Kerr metric}

The traditional form of MP equations is \cite{mathisson}%
\begin{equation}
\frac{D}{ds}\left(  mu^{\lambda}+u_{\lambda}\frac{DS^{\lambda\mu}}{ds}\right)
=-\frac{1}{2}u^{\pi}S^{\rho\sigma}R_{\pi\rho\sigma}^{\lambda}\label{mp1}%
\end{equation}

First of all, we consider the case of the motion of a spinning test particle
in equatorial circular orbits ($\theta=\pi/2$) from the weak Kerr source, that
is, $a/r\ll1$ and $MG/c^{2}$. For this case we take \cite{plyatsko 2013}%
\begin{equation}
u^{1}=0\text{, \ }u^{2}=0\text{, \ }u^{3}=\text{const, \ }u^{4}=\text{ const}%
\end{equation}
when the spin is perpendicular to this plane and the MP condition (\ref{mp}),
with%
\begin{equation}
S_{1}\equiv S_{r}=0\text{, \ }S_{2}\equiv S_{\theta}\neq0\text{, \ }%
S_{3}\equiv S_{\varphi}=0.
\end{equation}

The equation is given by%
\[
-y_{1}^{3}\ast y_{7}^{2}-2\alpha\ast y_{7}y_{8}+y_{8}^{2}-3\ast\alpha
\ast\varepsilon_{0}y_{7}^{2}+3\varepsilon_{0}y_{7}y_{8}-3\alpha\varepsilon
_{0}y_{8}^{2}y_{1}^{-2}+3\alpha\varepsilon_{0}y_{1}^{2}y_{7}^{4}%
\]%
\[
-\alpha\varepsilon_{0}\left(  1-\frac{2}{y_{1}}\right)  y_{8}^{4}y_{1}%
^{-3}+\varepsilon_{0}\left(  y_{1}^{6}-3y_{1}^{5}\right)  y_{7}^{3}y_{8}%
y_{1}^{-3}+\alpha\varepsilon_{0}\left(  3y_{1}^{3}-11y_{1}^{2}\right)
y_{7}^{2}y_{8}^{2}y_{1}^{-3}%
\]%
\begin{equation}
+\varepsilon_{0}\left(  -y_{1}^{3}+3y_{1}^{2}\right)  y_{7}y_{8}^{3}y_{1}%
^{-3}=0\label{c}%
\end{equation}

Then, for the case when the particle does not have spin the set of equations
(\ref{c}) with the dimensionless quantities \textit{y}$_{i}$ (\ref{y1}) and
(\ref{y5}) is reduced to%
\begin{equation}
-y_{1}^{3}\ast y_{7}^{2}-2\alpha\ast y_{7}y_{8}+y_{8}^{2}=0\label{a}%
\end{equation}
where $\alpha=a/M$.

In addition to Eq. (\ref{a}), we take the condition $u_{\mu}u^{\mu}=1$ and
obtain%
\begin{equation}
-y_{1}^{2}\ast y_{7}^{2}+4\alpha\frac{y_{7}y_{8}}{y_{1}}+\left(  1-\frac
{2M}{y_{1}}\right)  y_{8}^{2}=1\label{b}%
\end{equation}

We solve the system of equations (\ref{a}) and (\ref{b}) for the case of a
circular orbit and obtain the values of \ $y_{7}=Mu^{3}$ and\ $y_{8}=u^{4}$.

\subsection{Constants of motion for a weak Kerr metric}

With the Tulczyjew-Dixon condition (\ref{cond1}), we determine the center of
mass of the particle and let $u^{\mu}$ be its four-velocity; also, the MPD
equations (\ref{mov1}) and (\ref{mov2}) yield a unique $u^{\mu}$, namely
\cite{kyrian}%

\begin{equation}
u^{\mu}=V^{\mu}+\frac{1}{2}\left(  \frac{S^{\mu\nu}R_{\nu\rho\sigma\kappa
}V^{\rho}S^{\sigma\kappa}}{m^{2}+\frac{1}{4}R_{\chi\xi\zeta\eta}S^{\chi\xi
}S^{\zeta\eta}}\right)  .\label{condition}%
\end{equation}
$u^{\mu}$ and $V^{\mu}$ are named by Dixon as kinematical four velocity and
dynamical four velocity, respectively \cite{dixon 1979}.

Since $p_{\mu}p^{\mu}=$ constant and $S_{\rho\sigma}S^{\rho\sigma}=$ constant
along the particle trajectory \cite{wald}, we may set%

\begin{equation}
u_{\mu}u^{\mu}=-1\text{, \ \ \ \ \ \ }S_{0}^{2}=S_{\mu}S^{\mu}=\frac{1}%
{2m^{2}}S_{\mu\nu}S^{\mu\nu}\text{,}\label{7a}%
\end{equation}
and with these expressions, we obtain the center of mass condition and the
relation between the spin tensor and the vector spin.

Next, we reduced the set of MPD equations given by Plyatsko \textit{et al.
}\cite{plyatsko} for the case when a spinning test particle is a weak Kerr
metric in the equatorial plane and follows a circular orbit. Then, for the
initial conditions we need the values of both the energy ($E$) and the
component $z$ of the angular momentum ($J$) for a weak Kerr metric. In this
case, the constants of motion are given by%

\begin{align}
E  &  =m\left(  g_{44}+g_{34}\right)  V^{4}+\frac{Ma}{r^{2}}S^{13}-\frac
{M}{r^{2}}\frac{g_{33}V^{3}}{g_{44}V^{4}}S^{13}\label{E}\\
J_{z}  &  =-m\left(  g_{33}+g_{34}\right)  V^{3}+r\sin^{2}\theta S^{13}%
-\frac{Ma}{r^{2}}\frac{\left(  g_{33}+g_{34}\right)  V^{3}}{\left(
g_{44}+g_{34}\right)  V^{4}}S^{13}\label{J}%
\end{align}
where $V^{3}$ and $V^{4}$ are components of the dynamical 4-velocity and
$S^{13}$ is the perpendicular component of the spin vector.

We yield the components of the 4-velocity $u^{\lambda}$ (\ref{condition}) for
the case of a spinning test particle in a weak Kerr metric and in the
equatorial plane $\theta=\pi/2$ when spin is orthogonal to this plane and has
a constant radius ($x^{1}=r=$ constant). We use the Boyer-Lindquist
coordinates ($x^{1}=r,$ $x^{2}=\theta$, $x^{3}=\varphi$, $x^{4}=t$). We have%
\begin{align}
u^{1}  &  =0\text{, \ \ \ \ \ }u^{2}=0\text{, \ \ \ \ \ }u^{3}\neq0\text{,
\ \ \ \ \ }u^{4}\neq0\text{,}\label{29}\\
S^{12}  &  =0\text{, \ \ \ \ \ }S^{23}=0\text{, \ \ \ \ \ }S^{13}%
\neq0\label{30}%
\end{align}

In addition to (\ref{30}) by Tulczyjew-Dixon condition (\ref{cond1}) we write%
\begin{equation}
S^{14}=-\frac{P_{3}}{P_{4}}S^{13}\text{, \ \ \ \ \ }S^{24}=0\text{,
\ \ \ \ \ }S^{34}=\frac{P_{1}}{P_{4}}S^{13}\label{31}%
\end{equation}

Using (\ref{7a}), (\ref{29})-(\ref{31}) and taking the components of the
Riemann tensor for the weak Kerr metric in the equatorial plane, from
(\ref{condition}) we obtain%
\begin{align}
u^{1} &  =NV^{1}\left(  1+\frac{3M}{r^{3}}V_{3}V^{3}\frac{S_{0}^{2}}%
{m^{2}\Delta}+a\frac{S_{0}^{2}}{m^{2}\Delta}k_{1}\frac{V_{3}}{V_{4}}\right)
\nonumber\\
u^{2} &  =V^{2}=0\nonumber\\
u^{3} &  =NV^{3}\left(  1+\frac{3M}{r^{3}}\left(  V_{3}V^{3}-1\right)
\frac{S_{0}^{2}M}{m^{2}\Delta}+a\frac{S_{0}^{2}M}{m^{2}\Delta}k_{3}\frac
{V^{3}}{V^{4}}\right) \nonumber\\
u^{4} &  =NV^{4}\left(  1+\frac{3M}{r^{3}}V_{3}V^{3}\frac{S_{0}^{2}}%
{m^{2}\Delta}+a\frac{S_{0}^{2}M}{m^{2}\Delta}k_{4}\left(  \frac{V^{3}}{V^{4}%
}\right)  ^{2}\right) \label{32}%
\end{align}
where the constants $k_{1}$, $k_{3}$ and $k_{4}$ are given by%
\begin{align*}
k_{1} &  =\frac{3M\left(  1-\frac{4M}{3r}\right)  }{g_{11}g_{44}}\\
k_{3} &  =k_{1}\\
k_{4} &  =\frac{k_{1}}{rg_{44}}%
\end{align*}
and the expression $\Delta=1+\frac{1}{4m^{2}}R_{\chi\xi\zeta\eta}S^{\chi\xi
}S^{\zeta\eta}$\ for the weak Kerr metric is given by%
\begin{equation}
\Delta=1+\frac{S_{0}^{2}M}{m^{2}r^{3}}\left(  1-3V_{3}V_{3}-aA\frac{V^{3}%
}{V^{4}}\right) \label{33}%
\end{equation}
where $a=J/Mc$ is the angular density of the central mass and
\[
A=\frac{3M\left(  1-\frac{4M}{3r}\right)  g_{33}}{g_{44}}%
\]

We insert (\ref{33}) into (\ref{32}), we get%
\begin{align}
u^{1}  &  =\frac{NV^{1}}{\Delta}\left(  1+\frac{S_{0}^{2}M}{m^{2}r^{3}}%
-a\frac{S_{0}^{2}}{m^{2}}\left(  MA-k_{1}\right)  \frac{V^{3}}{V^{4}}\right)
\nonumber\\
u^{3}  &  =\frac{NV^{3}}{\Delta}\left(  1-\frac{2S_{0}^{2}M}{m^{2}r^{3}%
}-a\frac{S_{0}^{2}}{m^{2}}\left(  MA-k_{3}\right)  \frac{V^{3}}{V^{4}}\right)
\nonumber\\
u^{4}  &  =\frac{NV^{4}}{\Delta}\left(  1+\frac{2S_{0}^{2}M}{m^{2}r^{3}%
}-a\frac{S_{0}^{2}}{m^{2}}\left(  MA-k_{4}\left(  \frac{V^{3}}{V^{4}}\right)
^{2}\right)  \frac{V^{3}}{V^{4}}\right) \label{34}%
\end{align}

We introduce%
\begin{equation}
\varepsilon=\frac{\left\vert S_{0}\right\vert }{mr}%
\end{equation}
and obtain the expression for $N$ from the conditions%
\begin{align*}
V_{\lambda}V^{\lambda}  &  =0\\
V_{\lambda}S^{\lambda\nu}  &  =0
\end{align*}

$N$ is given by%
\begin{equation}
N=\frac{\Delta}{R}\label{36a}%
\end{equation}
where%
\begin{equation}
R=\left(  m^{2}\Delta^{2}+S_{0}^{4}R^{\mu\tau\rho\delta}R_{\mu\tau\rho\delta
}\right)  ^{%
\frac12
}\label{36}%
\end{equation}

We insert (\ref{36a}) into (\ref{34}) and obtain the components from
$V^{\lambda}$%
\begin{align}
V^{1}  &  =\frac{Ru^{1}}{\left(  1+\frac{S_{0}^{2}M}{m^{2}r^{3}}\left(
1+a\left(  A+k_{1}\right)  \right)  \right)  }\nonumber\\
V^{3}  &  =\frac{Ru^{3}}{\left(  1-\frac{S_{0}^{2}M}{m^{2}r^{3}}\left(
1+a\left(  A-k_{3}\right)  \right)  \right)  }\nonumber\\
V^{4}  &  =\frac{Ru^{4}}{\left(  1+\frac{S_{0}^{2}M}{m^{2}r^{3}}\left(
1-a\left(  A+3\left(  u^{3}\right)  ^{2}k_{4}\right)  \right)  \right)
}\label{37}%
\end{align}

We replace the components of the dynamical 4-velocity (\ref{37}) in the
constants of motion (\ref{E}) and (\ref{J}) for the case of a spinning test
particle in a weak Kerr field.

\section{Gravitomagnetic clock effect for spinning test particles}

For cheking our results, we review the papers in regarding to Gravitomagnetic
clock effect \cite{tsoubelis} and compare their numerical results with ours.
There is a phenomenon called the gravitomagnetic clock effect which consists
of a difference in the time it takes for two test particles to travel around a
rotating massive body in the equatorial plane and in opposite directions
\cite{faruque2004}. This difference is given by $t_{+}-t_{-}=4\pi a/c$, where
$a=J/Mc$ is the angular density of the central mass. Tartaglia has studied the
geometrical aspects of this phenomenon \cite{tartaglia}, \cite{tartaglia2001}
and Faruque yields the equation of the gravitomagnetic clock effect with spin
as%
\begin{equation}
t_{+}-t_{-}=4\pi a-6\pi S_{0}\text{,}\label{gm1}%
\end{equation}
where $S_{0}$ is the magnitude of the spin.

In true units this relation is given by%
\begin{equation}
t_{+}-t_{-}=\frac{4\pi J_{M}}{Mc^{2}}-\frac{6\pi J}{mc^{2}},\label{gm}%
\end{equation}
where the first relation of the right could be used to measure $J/M$ directly
for an astronomical body; in the case of the Earth $t_{+}-t_{-}\simeq10^{-7}%
\operatorname{s}%
$, while for the Sun $t_{+}-t_{-}\simeq10^{-5}%
\operatorname{s}%
$ \cite{mash 1999}.

\section{Numerical comparison for spinless and spinning test particle via MPD
equations}

In this section we give the numerical results for the case of a spinning
satellite orbiting around the Earth \cite{iorio 2001}. We took the data of the
Ariane-5 satellite which is a space European vehicle that is part of the
Ariane family \cite{ariane}. For our calculations we assume the satellite
follows a circular orbit with radius equal to $3.5\times10^{6}%
\operatorname{m}%
$ and travels in the equatorial plane. Of course, the satellite has an orbit
that is more complex. The initial conditions are given by geometrized units,
where the gravitational constant $G$, and speed of light $c$, are set equal to one.

According with the features of the Ariane-5 satellite, its initial conditions
are%
\[
\text{Mass}(m)_{\text{satellite}}=3\times10^{3}%
\operatorname{kg}%
\text{, \ \ \ \ \ \ \ }\varepsilon_{0}=\frac{S_{0}}{mr}=6.0976\times10^{-11}%
\]

The fundamental frequency in the longitudinal axis equals 30 Hz and the
components of the four velocity of the satellite are given by a set of
equations (\ref{c}) and (\ref{b}) for a orbit of $3.5\times10^{6}%
\operatorname{m}%
$. For this case, in ordinary units, the azimuthal component is $u^{3}%
=2.42294\times10^{3}$ $%
\operatorname{m}%
/%
\operatorname{s}%
$\ .

We take the set of MPD equations for a spinning test particle in a weak Kerr
metric (Appendix A) and write the initial conditions for this satellite. With
the Runge-Kutta method of order 4 \cite{press}, we obtain the cartesian
coordinates for a circular orbit when the satellite orbits in the same sense
of rotation of the central source ($a$). The program code is in the Appendix
B. We register the time that satellite takes for doing a lap. Then we take the
same data in the case when the satellite orbits with the sense of rotation
contrary to that of the central source ($a$). Finally, we take the difference
of time in these two orbits and obtain%
\begin{equation}
\Delta\tau_{\text{spinning}}=\tau_{+}-\tau_{-}=7.275957\times10^{-7}%
\operatorname{s}%
\end{equation}

Now we take the case when the test particle does not have spin and calculate
the cartesian coordinates ($x$, $y$, \thinspace$z$) for a circular orbit of a
spinless test particle around a rotating body mass both in the same sense of
rotation of the central mass and in opposite direction. In this case, we take
the set of MPD equations for a spinless particle, Eqs. (\ref{a}) - (\ref{b}).
There is a spinless particle in the equatorial plane ($u^{2}=0$) and with a
radius constant ($u^{1}=0$). We assume the same initial conditions as in the
previous case. As the above part, we calculate the difference of time of two
particles travel in the same orbit, but in opposite directions, and the result
is%
\begin{equation}
\Delta\tau_{\text{spinless}}=\tau_{+}-\tau_{-}=9.01062\times10^{-7}%
\operatorname{s}%
\end{equation}

This result is according to the literature. In some papers this difference of
time is called Effect Gravitomagnetic and is given by the expression
\cite{iorio 2001}%
\begin{equation}
\left(  \tau_{+}-\tau_{-}\right)  _{\phi=2\pi}\simeq4\pi\frac{J_{\oplus}%
}{M_{\oplus}c^{2}}\simeq10^{-7}%
\operatorname{s}%
\end{equation}
where $M_{\oplus}$ and $J_{\oplus}$ are the values of the mass of Earth and
the angular momentum respect

According to the results, the spinless test particle in a positive sense
completes a full orbit before the particle with the sense of rotation
contrary. This phenomenon is due to drag of the inertial frames with respect
to infinity and is called the Lense-Thirring effect \cite{mash 1984}. In the
case of the spinning test particles, not only there is a difference in the
time given by the Lense-Thirring effect, but also by a coupling between the
angular momentum of the central body with the spin of the particle
\cite{chandra}. The features change if the test particle rotates in one
direction or the other; therefore, the period is different for one sense and
for the other, and if the particle has spin or not. The difference of time
between the spinless particles and the spinning particles is so small that the
result is the same order of the shift ($10^{-7}%
\operatorname{s}%
$). In other words, when the spinning test particle is very small compared
with the central mass, the influence of the value of spin in the shift of time
is insignificant in regard to lapse of time.

\section{Conclusions}

In this paper, we take the Mathisson-Papapetrou-Dixon (MPD) equations given by
Plyatsko \textit{et al.} and obtained explicitly the MPD equations for the
case when the spinning test particle is orbiting in a rotating weak field.
Work that was not in the literature. In addition, we gave a scheme for the
eleven equations of the full set of equations of motion when the particle is
orbiting a rotating gravitational field. In the second part, we worked the
constants of motion such as the energy ($E$) and the angular momentum ($J_{z}%
$) of the spinning test particle in a weak Kerr metric. Finally, we calculated
the trajectories in cartesian coordinates ($x$, $y$, \thinspace$z$) of test
particles both spinless and spinning orbiting in a weak Kerr metric and
compared the time of two circular orbits in the equatorial plane for two test
particles that travel in the same orbit but in opposite directions. In the
case of the Earth, both for the spinless particles and the spinning particles
there is a difference of time in their trajectories when they describe a full
revolution with respect to an asymptotically inertial observer. This
phenomenum is called Gravitomagnetic Effect. From this situation, we concluded
that this shift, in the case of the spinless test particles, is given by the
angular momentum from the central source which drags the inertial systems in
the same sense of the rotation of the rotating massive body. For the case of
the spinning test particles, this time lapse is given not only by the angular
momentum from the central mass, but also by the couple between the angular
momentum from the massive rotating body and the parallel component of the spin
of the test particle. In the MPD equations, this couple is given by the
relationship between the components of the Riemman tensor ($R^{\mu}$
$_{\nu\rho\sigma} $) and the spin tensor ($S^{\rho\sigma}$).

In the future we will work in the set of Equations of motion of a test
particle both spinless and spinning for spherical orbits, that is, with
constant radius and out of the equatorial plane in a weak Kerr metric. In
addition, we are interesting in relating these equations with the experiments
type Michelson and Morley.

\begin{acknowledgement}
One of the authors is grateful with Pontificia Universidad Javeriana at
Bogot\'{a} and with Professor Roman Plyastsko for his helpful suggestions.
\end{acknowledgement}

\bigskip

\begin{center}
\textbf{Appendix A}

\textbf{Weak Kerr Metric}
\end{center}

The components of a weak Kerr Metric are given by%

\[
g_{\mu\nu}=\left(
\begin{array}
[c]{cccc}%
1-\frac{2M}{r} & 0 & 0 & \frac{2Ma\sin^{2}\theta}{r}\\
0 & -1-\frac{2M}{r} & 0 & 0\\
0 & 0 & -r^{2} & 0\\
\frac{2Ma\sin^{2}\theta}{r} & 0 & 0 & -r^{2}\sin^{2}\theta
\end{array}
\right)
\]

\bigskip

\begin{center}
\textbf{Appendix B}

\textbf{Program code}
\end{center}

$TF=1\ast10^{6};\epsilon_{0}=6.0976\ast10^{-11};y1=3.1085\ast10^{4}%
;y10=y1\ast\epsilon_{0};$

$y5=0;\alpha=1.9765\ast10^{-16};$

$M=4.431948\ast10^{-3};m=2.228\ast10^{-24};y2=\frac{\pi}{2};$

SetPrecision$[$NSolve[\{$-(y1)^{3}(y_{7})^{2}-2\ast\alpha\ast y_{7}%
y_{8}+(y_{8})^{2}-3\ast\alpha\ast\epsilon_{0}(y1)^{2}\ast\frac{(y_{7})^{2}%
}{(y1)^{2}}+3\ast\epsilon_{0}\ast y_{7}y_{8}-3\ast\alpha\ast\epsilon_{0}%
\ast\frac{(y_{8})^{2}}{(y1)^{2}}+$

$3\ast\alpha\ast\epsilon_{0}\ast(y1)^{2}(y_{7})^{4}-\alpha\ast\epsilon_{0}%
\ast(1-\frac{2}{y1})\frac{(y_{8})^{4}}{(y1)^{3}}+\epsilon_{0}((y1)^{6}%
-3(y1)^{5})\frac{(y_{7})^{3}y_{8}}{(y1)^{3}}+\alpha\ast\epsilon_{0}%
(3(y1)^{3}-11(y1)^{2})\ast\frac{(y_{7})^{2}(y_{8})^{2}}{(y1)^{3}}+$

$\epsilon_{0}\ast(-(y1)^{3}+3(y1)^{2})\frac{y_{7}(y_{8})^{3}}{(y1)^{3}%
}==0,-(y1)^{2}(y_{7})^{2}+\frac{4\ast\alpha\ast y_{7}\ast y_{8}}{y1}$

$+\left(  1-\frac{2}{y1}\right)  (y_{8})^{2}==1\},\{y_{7},y_{8}\}],10]$

system1 $=\{y_{3}%
\acute{}%
[s]==y_{7},y_{4}%
\acute{}%
[s]==y_{8},y_{3}[0]==0,y_{4}[0]==0\};$

$sol1=$ NDSolve$[$system1$,\{y_{3},y_{4},y_{3}%
\acute{}%
,y_{4}%
\acute{}%
\},\{s,0,TF\},$

$Method\rightarrow"Automatic",MaxSteps\rightarrow1\ast10^{10}]$

$graph1=$ ParametricPlot3D$[$Evaluate$[\{(y1)\ast\sin\frac{\pi}{2}\ast
\cos[{{{{{y_{3}[s]}}}}}],(y1)\ast\sin\frac{\pi}{2}\ast\sin[{{{{{y_{3}[s]}}}}%
}],(y1)\ast\cos\frac{\pi}{2}\}/.sol1],\{s,0,TF\},$

$AxesLabel->\{"x","y","z"\},PlotStyle->\{Blue\}]$

SetPrecision$[Table[\{s,y1\ast\sin[y2]\ast\sin[{{{{{y_{3}[s]}}}}%
}]\}/.sol1,\{s,0,TF,9.53674316406\ast10^{-7}\}],20]$

\end{document}